\documentclass[prx, letterpaper, superscriptaddress, amsmath, amssymb, amssymb, reprint, floatfix]{revtex4-2}

\usepackage{graphicx}
\usepackage{dcolumn}
\usepackage{bm}
\usepackage{xcolor} 
\usepackage[normalem]{ulem}

\begin{document}

\preprint{APS/123-QED}

\title{Multi-module microwave assembly for fast read-out and charge noise characterization of silicon quantum dots}

\author{Felix-Ekkehard von Horstig}
 \affiliation{Quantum Motion, 9 Sterling Way, London, N7 9HJ, United Kingdom}
 \email{fer28@cam.ac.uk}
 \affiliation{Department of Materials Sciences and Metallurgy, University of Cambridge, Charles Babbage Rd, Cambridge CB3 0FS, United Kingdom}

 \author{David J. Ibberson}
 \affiliation{Quantum Motion, 9 Sterling Way, London, N7 9HJ, United Kingdom}
 \author{Giovanni A. Oakes}
 \affiliation{Quantum Motion, 9 Sterling Way, London, N7 9HJ, United Kingdom}
 \affiliation{Cavendish Laboratory, University of Cambridge, J.J. Thomson Avenue, CB3 0HE, United Kingdom}
 \author{Laurence Cochrane}
 \affiliation{Quantum Motion, 9 Sterling Way, London, N7 9HJ, United Kingdom}
 \affiliation{Nanoscience Centre, Department of Engineering, University of Cambridge, JJ Thomson Avenue, CB3 0FF, United Kingdom}
 \author{David F. Wise}
 \affiliation{Quantum Motion, 9 Sterling Way, London, N7 9HJ, United Kingdom}
 \author{Nadia~Stelmashenko}
 \affiliation{Department of Materials Sciences and Metallurgy, University of Cambridge}
 \author{Sylvain Barraud}
 \affiliation{CEA, LETI, Minatec Campus, F-38054 Grenoble, France}
 \author{Jason~A.~W.~Robinson}
 \affiliation{Department of Materials Sciences and Metallurgy, University of Cambridge}
 \author{Frederico Martins}
 \affiliation{Hitachi Cambridge Laboratory, J.J. Thomson Avenue, CB3 0HE, United Kingdom}
 \author{M. Fernando Gonzalez-Zalba}
 \affiliation{Quantum Motion, 9 Sterling Way, London, N7 9HJ, United Kingdom}
 \email{fernando@quantummotion.tech}

\date{\today}

\begin{abstract}

Fast measurements of quantum devices is important in areas such as quantum sensing, quantum computing and nanodevice quality analysis. Here, we develop a superconductor-semiconductor multi-module microwave assembly to demonstrate charge state readout at the state-of-the-art. The assembly consist of a superconducting readout resonator interfaced to a silicon-on-insulator (SOI) chiplet containing quantum dots (QDs) in a high-$\kappa$ nanowire transistor. The superconducting chiplet contains resonant and coupling elements as well as $LC$ filters that, when interfaced with the silicon chip, result in a resonant frequency $f=2.12$~GHz, a loaded quality factor $Q=850$, and a resonator impedance $Z=470$~$\Omega$. Combined with the large gate lever arms of SOI technology, we achieve a minimum integration time for single and double QD transitions of 2.77~ns and 13.5~ns, respectively. We utilize the assembly to measure charge noise over 9 decades of frequency up to 500~kHz and find a 1/$f$ dependence across the whole frequency spectrum as well as a charge noise level of 4~$\mu$eV/$\sqrt{\text{Hz}}$ at 1~Hz. The modular microwave circuitry presented here can be directly utilized in conjunction with other quantum device to improve the readout performance as well as enable large bandwidth noise spectroscopy, all without the complexity of superconductor-semiconductor monolithic fabrication. 
\end{abstract}

\maketitle


\section{Introduction}

Fast charge readout is a desired quality for many solid-state quantum technologies~\cite{Petta2005,Persson2010a,Veldhorst2015,Hendrickx2021, Zhou2022}. It enables dynamical studies of these systems~\cite{Bluhm2011,Veldhorst2014}, provides a means to  explore their noise spectrum~\cite{Spietz2003} and for example, finds applications in qubit state readout~\cite{Connors2020,noiri2020,Oakes2023}, measurement of mechanical motion~\cite{Pearson2021} and fast thermometry at the nanoscale~\cite{Chawner2021,Blanchet2022}.

Particularly, for silicon-based spin qubits, fast charge measurements along with spin-to-charge conversion techniques~\cite{Ono2002,Elzerman2004} have enabled high-fidelity read out in short timescales, fulfilling one of the requirements for fault-tolerant quantum computing~\cite{Fowler2012}. In conjunction with high-fidelity one- and two-qubit gates~\cite{Xue2022,Noiri2022, Mills2022}, the demonstrations of simple instances of quantum error correction~\cite{Takeda2022}, the operation of a six-qubit processors~\cite{Philips2022} and the compatibility with advanced manufacturing techniques~\cite{Maurand2016,elsayed2022,Zwerver2022,Xue2021,Ruffino2022}, these steps show important progress towards a scalable silicon-based quantum computer~\cite{gonzalezzalba2021}. 

Silicon spin qubits are typically read using dissipative charge sensors like the single-electron transistor (SET) whose bandwidth can be increased to the MHz range by embedding the device in a high-frequency resonator~\cite{vigneau2022}.
To minimise the impact of the sensing method on quantum processor layout~\cite{Veldhorst2017,Vandersypen2017, Li2018, Crawford2023}, recent work has focused on resonant dispersive readout methods such as the single-electron box (SEB)~\cite{House2016,Urdampilleta2019} and in-situ readout \cite{Petersson2010,Colless2013}. Monolithic integration of the device and resonator on the same chip has lead to the highest charge readout rates both for the SEB~\cite{Borjans2021} as well as for in-situ readout~\cite{Zheng2019} due to the reduced parasitics that result in higher operation frequency, quality factor and resonator impedance. Additionally, monolithically integrated resonators have been used to explore circuit quantum electrodynamics with silicon spins~\cite{harvey-collard2022,mi2018,yu2023}. However, monolithic integration poses important fabrication challenges and may interfere with high qubit connectivity. On the other hand, multi-module microwave assemblies, in which the resonant circuitry is fabricated in a separate chiplet to the device~\cite{Holman2021}, offers the flexibility to independently optimize the layout and fabrication recipes for both technologies at the expense of typically lower performance. Recent advances on microwave assembly using inductively coupled superconducting resonators, however, have resulted in values at the state-of-the-art, specially for systems with high device-resonator coupling~\cite{Ibberson2021}.          

In this work, we develop a superconductor-semiconductor multi-module microwave assembly consisting of a Niobium Titanium Nitride (NbTiN) on sapphire superconducting chiplet connected to a fully-depleted silicon-on-insulator (SOI) high-$\kappa$ oxide transistor housing multiple quantum dots (QDs). The design is optimised to achieve a large internal quality factor $Q_{int}$ of 1600 and a resonant frequency $f_\text{r}$ of 2.1~GHz. 
We characterise the sensor sensitivity to dot-to-reservoir transitions (DRT) corresponding to the SEB mode of operation, and an interdot-charge transition (ICT) corresponding to an in-situ measurement. We find a minimum integration time [for a signal-to-noise ratio (SNR) of 1] $t_\text{min}$ of 2.77~$\pm$~0.01~ns and 13.5~$\pm$~0.3~ns for the SEB and in-situ measurements respectively, which come near to the state-of-the-art achieved using monolithic resonators and quantum-limited amplifiers. Furthermore, the performance of the SEB reported here approaches that of the best SET. 

Furthermore, we use the large sensitivity of the assembly to characterise interdot charge noise at high frequencies. We find a 1/$f$ dependence over nine decades of frequency up to 500~kHz and a charge noise level of 4~$\mu$eV/$\sqrt{\text{Hz}}$ at 1~Hz. Our result demonstrates a simple method to extend the high frequency end of the spectrum beyond those achieved using qubit measurements~\cite{Yoneda2017} and highlights that low charge noise levels can be achieved with high-$\kappa$ oxides. While our results are based on silicon devices, the microwave module presented here can be readily utilized with any charge-based system to facilitate improved readout performance.

\section{Resonator design}

\begin{figure}
\includegraphics{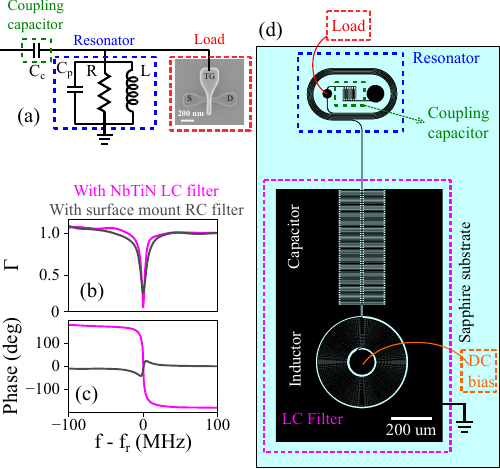}
\caption{\label{fig:Resonator_design}(a) A hanger mode $LCR$ resonator used for capacitive sensing consisting of the resonator (blue), a load (red) illustrated by an SEM image of a similar device} and a coupling capacitor (green) for impedance matching. The response of the load can be understood as an equivalent parallel Sisyphus resistance and parametric capacitance.
Reflected magnitude (b) and Phase (c) near the resonance frequency. Pink represents the resonator with the $LC$ filter while grey is the resonator bonded to an alternative surface mount $RC$ filter. (d) CAD design of the $LC$ resonator and filter used in this study. Black tracks represent NbTiN. Indicated are the resonator (blue), the coupling capacitor (green), the $LC$ filter (pink), and where the load and DC biases are wire-bonded to (red and orange). The entire resonator and $LC$ filter reside on the same Sapphire substrate.
\end{figure}

In Fig.~\ref{fig:Resonator_design}a, we present the equivalent circuit of the hanger mode resonator we use to read quantum devices~\cite{vigneau2022}. It consists of the resonator --- represented by an inductor ($L$) in parallel with a capacitance ($C_\text{p}$) and dissipative losses ($R$) --- coupled to an RF line via a coupling capacitor ($C_\text{c}$). The resonator is connected to a load (i.e. the quantum device) which, for QD devices, can be represented as the parallel combination of a variable capacitance (the quantum capacitance, $C_\text{Q}$) and the Sisyphus resistance ($R_\text{S}$)~\cite{Mizuta2017, Esterli2019}. 

We design the resonator based off previous multi-module approaches~\cite{Ahmed2018,Schaal2020,Ibberson2021} but make two significant improvements: (i) We utilise an on-chip coupling capacitor as well as a spiral inductor to enable independent control of the resonator frequency $f_\text{r}$, and the coupling coefficient, $\beta$. Further, the planar interdigitated capacitor minimises circuit losses with respect to surface mount components. (ii) We utilise superconducting $LC$ filters on the DC biasing line, instead of surface mount $RC$ filters. We integrate the $LC$ filters on the resonator chip. These reactive filters have been used in monolithically integrated resonators and increase the internal quality factor \cite{mi2017}, increasing the charge sensitivity but also reducing the size of coupling capacitor required for impedance matching. Off-chip lumped-element $LC$ circuits have been used in classical electronics using a flip-chip approach for filtering \cite{onizuka2007,takemura2013}, and as resonators to study the interaction of photons with spin ensembles \cite{huebl2013,zollitsch2015}. 
Additionally, flip-chip has been used to interface distributed microwave circuits (resonators and filters) with silicon QD systems~\cite{Holman2021,holman2020a}. 

We present the resonator implementation in Fig.~\ref{fig:Resonator_design}d. The resonator consists of a spiral inductor (blue dashed rectangle), surrounding an interdigitated coupling capacitor (green dashed rectangle). We utilize the spiral side of the capacitor to connect to devices (resulting in a parallel configuration with the resonator) while we bond the other side to the RF input line. We route the outer end of the inductor to an $LC$ filter (pink dashed rectangle) used to provide an RF ground while allowing DC signals to be applied. The $LC$ filter consists of an interdigitated capacitor to a ground plane on the sapphire chip, as well as a spiral inductor ($L_\text{filt} \sim$ 300 nH, $C_\text{filt} \sim$ 1 pF). All components are made of 80~nm NbTiN on Sapphire with a critical temperature of 12.5~ K and a kinetic inductance of 2.6 pH/$\square$ commercially fabricated by STAR Cryoelectronics. For specifics on the dimensions of the individual resonator component see Appendix~\ref{app:sec:resonator-layout}.\par

To quantify the improvement provided by the addition of the $LC$ filter to the resonator assembly, we connect the resonator by wire-bonding to a p-type FDSOI nanowire transistor of large dimensions (10~$\mu$m wide $\times$ 120~nm long $\times$ 8~nm high) to simulate a variable load impedance. For an image showing the wire-bonded setup, see Appendix \ref{app:sec:device-image}. The device has the same gate stack as the QD devices to be studied later: a 1.3 nm equivalent oxide thickness of high-$\kappa$ HfSiON gate dielectric and a combination of 5 nm thick TiN and 50 nm polycrystalline silicon layers as the gate metal. We then characterise the resonance using a vector network analyser (VNA) inside a pumped helium system at 2 K. We measure the resonator assembly in two different configurations and compare the results: (i) DC bias applied through the $LC$ filter (orange dashed square) and (ii) DC bias applied through a surface mount $RC$ low pass filter ($C=100$~pF and $R=10$ k$\Omega$) bonded to the load side of the inductor (effectively bypassing the $LC$ filter). We plot the magnitude ($\Gamma$) and phase of the reflection coefficient in the two resonator configurations in Fig. \ref{fig:Resonator_design}b and c (pink and blue traces, respectively). From the plots, we extract the loaded and internal quality factors for case i(ii) of $Q=394(181)$ and $Q_{int}=900(280)$, respectively. We observe an improvement of a factor of $\sim$ 3 indicating the positive impact of the $LC$ filter in reducing internal losses. Furthermore, although not represented in the plot, we note a higher resonance with the $LC$ filter (2059~MHz vs 1880~MHz) which can be attributed to the filter capacitance $C_\text{f}$ resulting in a reduction in the effective inductance of the resonator, $L_{\text{eff}} \sim L [1-(C_\text{p}+C_\text{c})/C_\text{f})]$. For a detailed derivation see Appendix~\ref{sec:app:Cfilt-correction}. As we shall see in the later part of this Article, we observe larger quality factors when testing the assembly with smaller transistors at millikelvin temperatures. Such difference can be explained by (i) the larger area, and hence loss, of the transistors used in this preliminary study and (ii) the TiN layer of the gate stack turning superconducting at $\approx 1$~K.  

\section{Dispersive Interaction}

\begin{figure}
\includegraphics{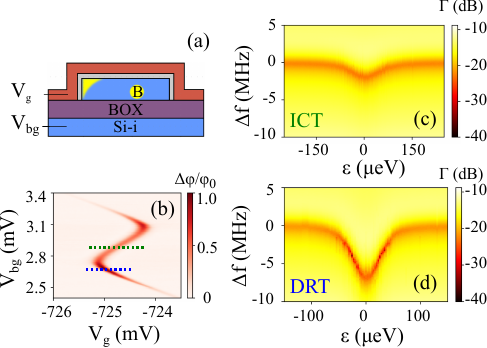}
\caption{\label{fig:Device-freqshift}(a) Cross section of the silicon nanowire transistor along the direction of the gate (not to scale), showing the TiN+Poly gate with HfSiON/SiO$_2$ dielectric surrounding a B-doped Si channel. The nanowire sits on a buried oxide (BOX) below which the silicon substrate can be used as a back-gate when illuminated. (b) Stability map indicating the location of the ICT (green) and DRT (blue) read-out locations. Frequency shifts vs detuning maps for the ICT (c) and DRT (d), respectively.}
\end{figure}

In this Section, we explore the dispersive interaction between the microwave resonator and QDs in a silicon nanowire transistor, see Fig. \ref{fig:Device-freqshift}a.  
We select a resonator with an effective $L$ of 31~nH, a $C_\text{c}$ of 40~fF and a $C_\text{p}$ (including that of the silicon chiplet) of 140~fF resulting in a resonance frequency of 2116~MHz and a resonator impedance $Z_\text{r} = 470$~$\Omega$. We connect the resonator to a single gate SOI nanowire transistor with a channel width of 120~nm, a length of 60~nm and height of 8~nm, and a Boron channel doping density of $5\cdot 10^{17}$~ cm$^{-3}$. We use additional $LC$ filters (part superconducting chiplet) on the source and drain electrodes.
We use the back-gate voltage $V_\text{bg}$ as well as the top-gate voltage $V_\text{g}$ to bias the device to an ICT between a corner dot~\cite{Voisin2014, Ibberson2018} and a Boron acceptor with overall charge configurations (N+1$_D$, 0$_B$) to (N$_D$,1$_B$), see Fig.~\ref{fig:Device-freqshift}b. Here, we plot the normalised resonator phase shift. Besides, we see signal at the DRT of the corner dot. We choose this voltage region for its low charge noise.

We characterise the ICT and DRT at the measurement points indicated by the dashed lines and find gate lever arms $\alpha_\text{ICT}=0.53 \pm 0.03$ and $\alpha_\text{DRT}=0.72 \pm 0.03$, extracted from Landau-Zener-St{\"u}ckelberg interferometry~\cite{Shevchenko2010,Gonzalez-Zalba2016}. 
We extract the tunnel coupling and tunnel rate of the ICT and DRT, respectively from their transition linewidth~\cite{Mizuta2017, Ahmed2018b} and obtain $2t_\text{c}/2\pi$ = 15.6 $\pm$ 0.4 GHz for the ICT and $\Gamma/2\pi$ = 10.2 $\pm$ 0.2 GHz for the DRT.

\par
Next, in Fig.~\ref{fig:Device-freqshift}c,d, we measure the resonator's dispersive shift induced by the ICT and DRT. To vary the detuning from the charge transition point, we change the top gate voltage. We fit the resonance at each point in detuning to extract the resonant frequency $f_\text{r}$ and line-width, $\kappa$. Away from the charge transitions, we extract a bare resonator frequency of $f_0=2.116$~GHz with linewidth $\kappa/2\pi = 2.5$~MHz ($\kappa_\text{int}/2\pi = 1.05$~MHz). We find a frequency shift of 1.8~MHz for the ICT and 5.6~MHz for the DRT. Further, for the ICT, we extract a coherent coupling rate $g_0/2\pi$=168~MHz, a charge decoherence rate of $\gamma/2\pi$=1.45~GHz and a tunnel coupling $2t_c/2\pi$=17.7~GHz
that closely matches the value extracted from lineshape analysis. Charge decoherence rates, $\gamma/2\pi$, in the range 1-10 GHz have previously been reported in silicon nanowire transistors like the one studied here \cite{dupont-ferrier2013, Gonzalez-Zalba2016, Anasua2018,Ibberson2021}.  For more details on the extraction of parameters from these fits, see Appendix~\ref{sec:app:fitting-freq-shift}. 

In both cases, $\Delta f$ is of the order of $\kappa$ which is the condition for maximum state visibility ($\Delta f/\kappa/2\pi=0.72$ for the ICT and 2.24 for the DRT)~\cite{Ibberson2021}. We exploit this regime for fast readout.

\section{MEASUREMENT TIME: DRT and ICT}

We now characterise the dispersive readout performance on the DRT and the ICT. To quantify the performance, we utilize the power SNR of the transitions at a given integration time, $t_\text{meas}$. The measurements at the DRT represent the expected charge readout performance of a SEB~\cite{House2016,CirianoTejel2021,Neigemann2022,Oakes2023,Hogg2023} while the measurements at the ICT are equivalent to the expected charge readout performance in Pauli-spin blockade-based schemes~\cite{Petersson2012,Betz2015,House2015,Pakkiam2018, West2019}.

\begin{figure}
\includegraphics{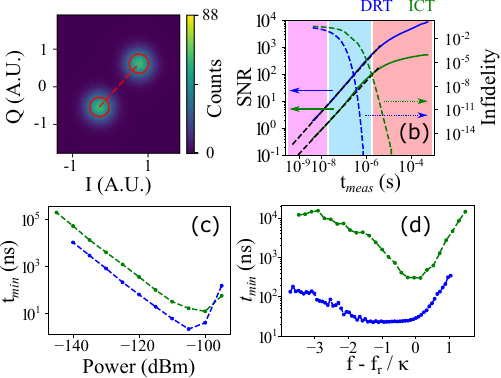}
\caption{\label{fig:SNR-tmin}(a) Histogram of signal and background measurements obtained for the DRT at $t_\text{meas}$ = 100~ns ($f=2116$~MHz, power of -105 dBm). In red, we show the distance and standard deviation of the two measurement outcomes used for the SNR calculations. (b) SNR and Infidelity as a function of $t_\text{meas}$ for the DRT (blue) and ICT (green) at the optimum read-out power of -105~dBm and frequency of 2116~MHz for the DRT and 2116.8~MHz for the ICT. The SNR is limited in the three regimes by finite bandwidth $t_0$ (pink), white noise (cyan) and 1/$f^\eta$ noise (red). (c) $t_\text{min}$ measured against power at frequencies of 2117.2~MHz for the ICT and 2115.4~MHz for the DRT. (d) $t_\text{min}$ against measurement frequency, measured at -120~dBm and -110~dBm for the ICT and DRT, respectively.}
\end{figure}

To extract the signal, we measure the difference in the IQ resonator response on and off the DRT (or ICT) peak. We first tune the device on (off) the measurement point, followed by a wait period (100 $\mu$s) much longer than the bandwidth of the DC line to stabilise the gate voltage and pre-condition the resonator, removing effects due to resonator ring-up. We then integrate the response for $t=t_\text{meas}$. We present the results in Fig.~\ref{fig:SNR-tmin}a where we plot a 2D histogram of the outcome of 1000 measurements where the IQ signal has been integrated for 100~ns. We observe two well-defined Fresnel lollipops with 2D standard deviation $\sigma_\text{on(off)}$ which allows defining the SNR,   

\begin{equation}\label{eq:SNR_equation}
    \text{SNR} = \frac{(I_\text{on}-I_\text{off})^2+(Q_\text{on}-Q_\text{off})^2}{0.25 (\sigma_\text{on} +\sigma_\text{off})^2}.
\end{equation}

We note that at low integration times, as in panel (a), the noise level for both states is equal and set by the cryogenic amplifier. Next, in Fig.~\ref{fig:SNR-tmin}b, we explore the dependence of the SNR with $t_\text{meas}$. To do this, we measure the signal using a sampling rate of 250 MHz (4 ns per point) with a total measurement time of 4 $\mu$s, to measure the low end of the $t_\text{meas}$ range, followed by a second measurement with a sampling rate of 5 MHz and total measurement time of 500 $\mu$s that covers the large $t_\text{meas}$ range. We then use a moving average to generate data at various effective $t_\text{meas}$. Each data point is composed of 1,000 shots on and off the ICT and DRT. We see that at low $t_\text{meas}$, the SNR follows the expected linear relationship for white noise-dominated data set by \cite{barthel2010}

\begin{equation}\label{eq:SNR_t0mod}
    \text{SNR} =  \left( \frac{t_\text{meas}+ t_0}{t_\text{min}}\right)^\zeta, 
\end{equation}

\noindent where $t_0$ relates to the bandwidth of the measurement set-up and $t_\text{min}$ corresponds to the integration time for SNR~$~=~1$ (at large measurement bandwidth) which we used as a characteristic figure of merit~\cite{vigneau2022}. Besides, $\zeta$ is determined by the noise type with $\zeta$~=~1 corresponding to white noise and $\zeta$~$<$~1 for pink and red noise.\par

At high $t_\text{meas}$ (red regime Fig.~\ref{fig:SNR-tmin}b), the data deviates from the linear trend and the SNR increases at a lower rate, characterised by $\zeta$~$<$~1. This can be explained by a transition from white noise to 1/$f^\eta$ noise-dominated spectrum with a corner frequency around 1~MHz, see Appendix \ref{sec:app:Fresnel-chargenoise} showing the distortion of the on-state Fresnel lollipop due to excess noise. At very low $t_\text{meas}$ (pink regime Fig.~\ref{fig:SNR-tmin}b), we also find a deviation from the linear trend which can be attributed to a finite bandwidth of the measurement setup on the order of $t_0$ $\sim$ 2~ns, matching well the bandwidth of the IQ demodulator at 275~MHz. At the fastest readout time achievable by our measurement setup (4 ns) we find an SNR of 1.88 (0.32) for the DRT (ICT).

By fitting the low $t_\text{meas}$ regime to Eq.~\ref{eq:SNR_t0mod} with $\zeta=1$, 
we extract a $t_\text{min}=13.5 \pm 0.3$~ns for the ICT and $t_\text{min}=2.77 \pm 0.01$~ns for the DRT.
To our knowledge, the DRT readout figure represents the lowest reported using a multi-module assembly and approaches the best reported values for monolithic resonator integration (1.8~ns~\cite{Borjans2021}) and SET readout (0.625~ns~\cite{Keith2019}). The ICT figure also comes near to those reported in the literature \cite{Ibberson2021,Stehlik2015}.

Additionally, in panel b, we plot the measurement infidelity, i.e. the likelihood of mislabelling the state of the system. The electrical infidelity is directly related to the SNR via,

\begin{equation}\label{eq:Fidelity-from-SNR}
    I_\text{E} = 1 - F_\text{E} = \frac{1}{2} \left[1 - \text{erf}\left(\frac{\text{SNR}}{2\sqrt{2}}\right)\right].
\end{equation}
\noindent where $F_\text{E}$ and $I_\text{E}$ are the electrical fidelity and infidelity respectively and erf is the error function. 
We achieve an electrical infidelity of $10^{-4}$ at 53 $\pm$ 7 (280 $\pm$ 70) ns for the DRT(ICT). Similarly, we achieve an infidelity of $10^{-8}$ at 0.12 $\pm$ 0.03 (0.6 $\pm$ 0.2) $\mu$s for the DRT(ICT).
We stress that this is the electrical infidelity corresponding to electrical readout errors associated to the quality of the sensor only. When used for qubit state read-out, state decay ($T_1$ processes) can also introduce readout errors~\cite{Laucht2021}. In any case, for a given $T_1$, a shorter measurement time reduces the likelihood of state decay and therefore improves fidelity. \par

\par
Using $t_\text{min}$ as a figure of merit for comparison~\cite{vigneau2022}, we now characterise the read-out performance against measurement power and frequency (Fig.~\ref{fig:SNR-tmin}c and d). For the measurement power, we find $t_\text{min}$ decreases for increasing power until it reaches a minimum at -100~dBm and -105~dBm for the ICT and DRT, respectively. This trend can be explained by noting that the signal strength is proportional to the number of photons populating the resonator. However, at intermediate powers, photon back-action causes a broadening of the line-width which reduces the dispersive frequency shift on the resonator, reducing the difference between the on and off states~\cite{Maman2020}. Finally, at high powers, the large driving induces increased kinetic inductance and losses in the resonator, decreasing its quality factor and therefore reducing its sensitivity~(see. Appendix \ref{sec:app:highpower-loss_lkin}).\par

Next, in Fig. \ref{fig:SNR-tmin}d, we measure $t_\text{min}$ as a function of frequency and fixed power (below power broadening levels). Under these conditions, the frequency shift caused by the charge transition is that shown in Fig. \ref{fig:Device-freqshift}c. We find that $t_\text{min}$ exhibits a plateau, where its value is minimum, within the frequency range between the bare and loaded frequency of the resonator, a frequency bandwidth of the order of $\Delta f$.
We note that when applying higher powers, close to the optimal, $\Delta f$ decreases and therefore narrows the optimal frequency window. The optimum read-out frequency for high power narrows down around the bare resonance frequency (\textcolor{red}{Cochrane et al. in preparation}).

\section{Charge noise measurements}
\begin{figure*}
\includegraphics{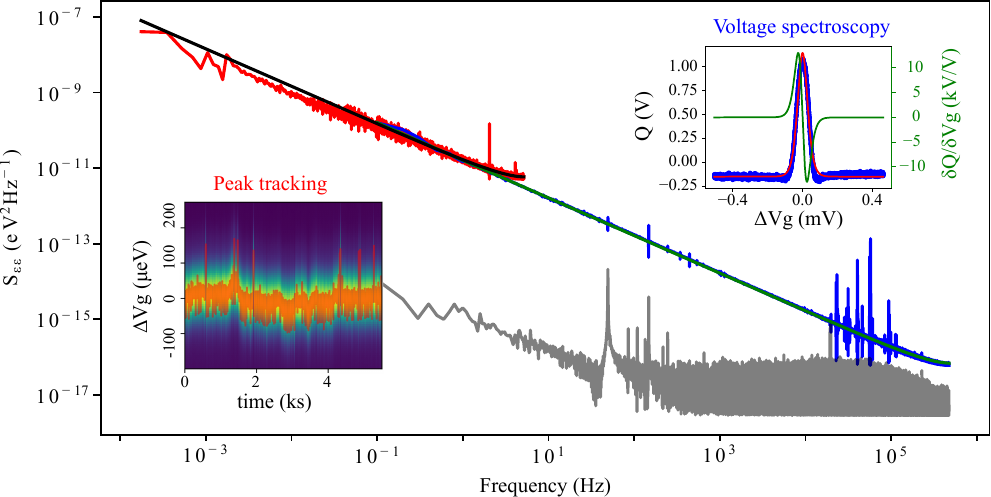}
\caption{\label{fig:Charge_noise} Charge noise spectral density against frequency from peak tracking (red) and voltage spectroscopy (blue). Fits are in black and green, respectively. The data in grey shows the equivalent setup noise.
Insert lower left: peak location as a function of time used to extract the charge noise. Insert upper right: Line trace of the ICT used to extract the slope of the signal with gate voltage.}
\end{figure*}

Having characterised the large SNR of the resonator, we now utilize it to determine the charge noise spectrum of the device. Particularly, we study charge noise at the ICT. Charge noise results in time-dependent variations of both the detuning and tunnel coupling between QDs impacting the fidelity of qubit operations~\cite{Yoneda2017}. Understanding its magnitude and frequency dependence is hence of high relevance to determine device quality. Charge noise is typically characterised in terms of the detuning noise spectral density ($S_{\epsilon\epsilon}$) which follows a power law dependence given by    

\begin{equation}\label{eq:See-bare}
    S_{\epsilon\epsilon} = \frac{S_0 }{f^\eta},
\end{equation}

\noindent where $\sqrt{S_0}$ is the characteristic charge noise at 1 Hz and the exponent $\eta$ is typically around 1 with values ranging between 0.5 and 2~\cite{Struck2020}. The power law dependence arises by considering an ensemble of two-level fluctuators (TLFs) with a uniform distribution of activation energies and interactions between them. Typically, charge noise is measured in the frequency range around 1 Hz and quoted in terms of the characteristic value $\sqrt{S_0}$. However, measuring charge noise at high frequencies is essential to determine whether the power law observed at low frequencies extends to the regime at which qubit operations occur. Here, we utilize the resonator to cover charge noise over 9 decades of frequency extending up to 500 kHz, see Fig~\ref{fig:Charge_noise}. To cover the frequency range, we utilize two methods: peak tracking and voltage spectroscopy~\cite{Kranz2020}.   

Peak tracking (bottom left insert) operates by repeatedly measuring the voltage at which the peak of the charge transition occurs. The voltage noise spectral density ($S_{VV}$) is then translated into the detuning noise spectral density by using the ICT lever arm: $S_{\epsilon\epsilon } = \alpha_\text{ICT}^2 S_{VV}$. We use peak tracking to cover the range $10^{-4}$ to $10^1$~Hz. Voltage spectroscopy (top right insert) quantifies the voltage signal fluctuations at a fixed detuning point, in this case the signal in quadrature ($Q$) at the point of highest derivative. The detuning noise spectral density can then be calculated as $S_{\epsilon\epsilon} = \alpha_\text{ICT}^2 S_{QQ}/(\partial Q/\partial V_\text{g})^2$ where $S_{QQ}$ is the quadrature voltage noise spectral density. We use voltage spectroscopy in the regime from $10^0$ to $10^5$~Hz. \par

Finally, we measure the resonator and measurement setup noise in the same way as voltage spectroscopy, however this time we detune the device away from any charge transitions. In order to compare this to the charge noise data we transform the data using the same conversion to report an equivalent $S_{\epsilon\epsilon}$ (see grey trace in Fig. \ref{fig:Charge_noise}). We find this noise to be well below the recorded $S_{\epsilon\epsilon}$ for all but the highest frequencies.

We fit Eq.~\ref{eq:See-corrected} to the peak tracking and voltage-spectroscopy data. We include a functional modification to the equation to account for aliasing due to finite sampling rate that becomes apparent as an upward bend at the high frequency end of both datasets~(See Appendix~\ref{sec:app:Chargenoise-sampling}). We find a charge noise of 3.9(4.0)~$\mu$eV/$\sqrt{\text{Hz}}$ and exponent of 0.99(0.99) for the peak tracking (voltage spectroscopy) method. In the low frequency regime of the peak-tracking below $10^{-2}$ Hz the charge noise drops off which may be due to the dominant effect of one or more TLFs with characteristic frequency below $10^{-2}$ Hz. Similar results have been reported in the literature \cite{elsayed2022,rudolph2019,spence2022} although further investigation would be necessary to confirm it in this device. We recall that the measurement location was partially chosen due to its low charge noise. We additionally measure charge noise at other combinations of gate and back-gate voltages of the device (not shown) and find $\sqrt{S_0}$ values in the range of 4.0~-~10~$\mu$eV/$\sqrt{\text{Hz}}$.

The charge noise figure demonstrated here is larger than the state-of-the-art for an MOS structure (0.61~$\mu$eV$/\sqrt{\text{Hz}}$~\cite{elsayed2022}) but is within the values reported for other SiO$_2$-only oxide nanowire transistors and  below the figure in the few-electron limit~\cite{spence2022}. We highlight that the oxide in our sample is composed of 0.8~nm SiO$_2$ and 1.9~nm HfSiON. Such high-$\kappa$ dielectrics have been associated with a high oxide trap density which could negatively impact the charge noise figure~\cite{Ibberson2018}. Our result demonstrates that high-$\kappa$ oxides are not necessarily worse than SiO$_2$-based devices while they provide tighter control of the QD electrostatics manifested in the large gate lever arms reported for high-$\kappa$ technology~\cite{Gonzalez-Zalba2015,Ahmed2018}.  

With the high SNR demonstrated here and the combination of the two methods, we cover the charge noise spectrum over 9 decades of frequency up to 500~kHz, an upper frequency range comparable to qubit operation rates. Previous work measured charge noise over 5 decades and up to 1~Hz using a combination of current spectroscopy, peak-tracking and variations in charge dephasing exchange control experiments~\cite{Kranz2020}. Struck et al.~\cite{Struck2020} measured the charge noise of an SET over 9 decades of frequency up to 10~kHz using current spectroscopy. Finally, CPMG methods have been used to study the impact of charge noise on qubit properties for frequencies up to 320~kHz~\cite{Yoneda2017} and 2~MHz~\cite{Connors2022}. Overall, our work demonstrates an easy to use method to study charge noise at the important high frequency end which is relevant to individual qubit operations and directly at the ICT without the need of separate charge sensors that may contribute to the charge noise figure. Besides, our approach may facilitate device quality characterisation and setup noise debugging at high frequencies without the need to resort to qubit measurements. 

\section{Discussion}

In this Article, we have achieved state-of-the-art charge readout rates utilizing a multi-module microwave assembly. To facilitate the discussion of our results, we will introduce the dependence of $t_\text{min}$ with system variables as calculated for small dispersive signals and large tunnel rates~\cite{Oakes2023},

\begin{equation}\label{eq:tminSEB_Oakes}
    t_\text{min} \propto \frac{(1+\beta)^2}{\beta} \frac{k_\text{B}T_\text{n}}{(e \alpha)^2} \frac{1}{Q_{int} Z_\text{r}} \frac{1}{f_\text{r}^2} 
\end{equation}
\noindent where $T_\text{n}$ is the noise temperature.

First of all, with the capacitively coupled microwave assembly, we have looked to comparatively increase the resonator frequency $f_\text{r}$ over other multi-module approaches while keeping control over the external coupling, $\beta$. Secondly, we have introduced an integrated $LC$ filter on the gate line to increase the quality factor of the resonator and obtained an improvement in $Q$ $(Q_{int})$ by a factor of 2.2 (3.2). Our data follows a similar trend to results on high frequency (6-8~GHz), high-impedance ($\sim$ k$\Omega$) distributed wave-guide resonators \cite{mi2017,Harvey-Collard2020} that showed an improvement of one order of magnitude. To understand the lower improvement factor, we refer to the expression of the microwave photon leak rate through the DC lines ($\kappa_{DC}$) \cite{mazin2005,Harvey-Collard2020}: $\kappa_{DC} = 16\pi^2 f_\text{r}^3 Z_\text{r} Z_\text{DC} C_\text{DC}^2$, where $Z_\text{DC}$ the impedance of the DC line and $C_\text{DC}$ is the capacitance from the resonator to the DC line. The aim of the $LC$ filter has been to reduce the $Z_\text{DC}$ a microwave photon experiences towards the DC line. The comparatively lower improvement in our case could be attributed to the lower $f_\text{r}$ and $Z_\text{r}$, making $\kappa_{DC}$ a less prominent loss mechanism.

To benchmark our results against the reports in literature, we use the figure of merit of $t_\text{min}$.  We note that the metric is load aware due to the lever arm, $\alpha$ (see Eq.~\ref{eq:tminSEB_Oakes}). Across the field, there are large differences in $\alpha$ ranging from values close to $\alpha\approx 1$, particularly thin SOI devices~\cite{Ahmed2018}, to much smaller values $\alpha\sim 0.1$, for example, in Si/SiGe quantum wells~\cite{Philips2022}. To facilitate a resonator-only comparison, we suggest an additional load-agnostic metric: $t_\alpha=t_{min}\alpha^2$. This choice of metric has two benefits: (i) It allows a better comparison of read-out performance of different resonator across different platforms and (ii) it gives a tool to estimate how well a particular superconducting chiplet would perform on a different quantum system. In Table \ref{tab:tmin-overview}, we show a non-exhaustive list of reflectometry based read-out results of QD systems extracted from the literature.\par  
 
 \begin{table*}
     \centering
     \begin{tabular}{l|c|c|c|c|c|c}
          Reference                             & Platform          & Sensor type  & Resonator type & $t_\text{min}$ (ns)    & $\alpha$  & $t_\text{min} \alpha^2$ (ns) \\ \hline
          \textbf{This work}                    & \textbf{Si-SOI nanowire}   & \textbf{SEB}  & \textbf{Multi-module} & \textbf{2.77}   & \textbf{0.72}    &  \textbf{1.44} \\
                                                     &                   & \textbf{In-situ}          & \textbf{Multi-module} & \textbf{13.5}    & \textbf{0.53}    &  \textbf{3.79} \\
        Borjans et al. \cite{Borjans2021}            & Si/SiGe           & SEB              & On-chip  & 1.8    & 0.1$^x$   &  0.018$^x$ \\
        Oakes, Ciriano-Tejel et al. \cite{Oakes2023} & Si-SOI nanowire  & SEB (JPA on)      & Multi-module & 100    & 0.35 & 12    \\
                                                     &                  & SEB (JPA off)     & Multi-module & 450    & 0.35 &  55  \\
                                                     & Si-SOI nanowire  & SEB               & Multi-module & 170    & 0.4  & 27  \\
          House et al. \cite{House2016}              & P donor is Si    & SEB               & Multi-module & 550    & 0.45 &  111      \\
          Hogg et al. \cite{Hogg2023}                & P donor is Si    & SEB               & Multi-module & 720    & 0.38 &  104   \\
          Stehlik et al. \cite{Stehlik2015}          & InAs nanowire   & In-situ (JPA on)   & On-chip  & 7      & 0.15* & 0.16*  \\
                                                     &                 & In-situ (JPA off)  & On-chip  & 16000  & 0.15* & 360*  \\
          Ibberson et al. \cite{Ibberson2021}        & Si-SOI nanowire & In-situ            & Multi-module & 10     & 0.72  & 5.2 \\

          Schaal et al. \cite{Schaal2020}            & Si-SOI nanowire & In-situ (JPA on)   & Multi-module & 80     & 0.36  &  10.4              \\
                                                     &                 & In-situ (JPA off)  & Multi-module & 1200   & 0.36  &  156 \\
          
          Zheng et al. \cite{Zheng2019}             & Si/SiGe          & In-situ            &  On-chip & 170    & Not given & -  \\
          Keith et al. \cite{Keith2019}             & P donor is Si    & SET                & Multi-module  & 0.625$^\dagger$ & NA & - \\
          Noiri et al. \cite{noiri2020}             & Si/SiGe          & SET                & Multi-module  & 22 $^\dagger$   & NA      &  - \\
          House et al. \cite{House2016}            & P donor is Si     & SET                & Multi-module  & 55              & NA      &  - \\
     \end{tabular}
     \caption{Minimum measurement time comparison of reported charge sensors in the literature. $^\dagger$ $t_\text{min}$ was not reported but the amplitude SNR was reported at a different integration time and $t_\text{min}$ was calculated assuming SNR$ = (t_\text{meas}/t_\text{min})^{1/2}$. *$\alpha$ was not reported but as an estimate $\alpha$ from a similar system and the same group but different paper was used \cite{jung2012}. $^x$ $\alpha$ not directly given in the paper but calculated from the FWHM and $\Gamma$ supplied.}
     \label{tab:tmin-overview}
 \end{table*}

First of all, we see that both the SEB and in-situ readout demonstrations in this Article perform near the state-of-the-art in terms of $t_\text{min}$. Besides, we note that, in our work, $t_\alpha$ at the DRT and the ICT differ by a factor of about 2.6. This difference, not captured in Eq.~\ref{eq:tminSEB_Oakes}, could be attributed to the increase in photon loss rate at the DRT ($\Delta \kappa$ = 1.8 MHz) compared to the ICT ($\Delta \kappa$ = 0.15 MHz), which can be understood as a resistive component to the signal, adding on top of the capacitive one.  

While both the SEB and in-situ readout results display $t_\text{min}$ approaching the state-of-the-art, a comparison through $t_\alpha$ shows that on-chip microwave resonators are ultimately a better choice in terms of enabling high measurement rates for a given technology. This can be particularly appreciated in the SEB results in SiGe~\cite{Borjans2021}. The confined geometry, use of high quality substrates and the use of high kinetic inductance materials in these systems enables a higher frequency, $Q$ and impedance.

However, the impact of the footprint of on-chip resonators have on quantum processor topology is not negligible and they come with the additional complexity of hybrid superconductor-semiconductor fabrication.

There is significant room for improvement for multi-module microwave assemblies.
High kinetic inductance materials such as Titanium Nitride \cite{yu2021} and granular Aluminium \cite{grunhaupt2018} could be used to further reduce the footprint of the resonators, reducing the parasitic capacitance and therefore increasing their impedance and frequency.

Additionally, the use of a Josephson Parametric Amplifier (JPA) could further improve readout of the microwave-assembly studied here. For example, we note that in-situ readout results with on-chip resonators have only reported a higher performance when in conjunction with a Josephson Parametric Amplifier (JPA) which has the role of reducing $T_\text{n}$~\cite{Stehlik2015}. In this context, a JPA in conjunction with the multi-module microwave assembly presented here could result in a reduction factor of $\approx$ 20 in $t_\text{min}$, given the typical noise temperature of state-of-the-art cryogenic amplifiers ($T_\text{n}=2.5~K$) and the frequency and temperature ($T=100$~mK) of operation of our system.     

Finally, we also benchmark our results against the most sensitive charge sensor to date: the rf-SET. Particularly, we see that the SEB demonstrated here ($t_{min}=2.77$~ns) approaches the best result for the SET, 625~ps~\cite{Keith2019}, supporting the idea that dispersive readout methods could be as sensitive as dissipative ones~\cite{Oakes2023}. In more detail, it was recently theoretically proposed that the signal generated by the single-electron AC currents in the SEB, when driven at high frequencies, could become of the order of the large single-electron currents flowing through SETs, resulting in comparable sensitivities under the same system noise levels. Furthermore, it was proposed that JPAs could reduce the noise levels more efficiently in dispersive systems given that the ultimate noise source, the Sisyphus noise~\cite{cochrane2022}, can be made smaller that the SET's shot noise. The results in this Article support this theory and encourage further measurements combining high-frequency strongly-coupled SEBs in conjunction with quantum-limited amplification. We also note that progress in dispersive readout methods over SETs would benefit the development of compact quantum processor topologies. In particular, dissipative charge sensors like the SET require three physical contacts additional to the qubit to be sensed, whereas the SEB just requires two. In the case of in-situ readout only the two electrodes controlling a double QD system are needed to implement the method.     

\section{Conclusion}

In conclusion, we have shown the design of a multi-module resonance circuit containing an integrated capacitive coupling element, a spiral inductor and $LC$ low-pass filters to the bias lines.  We use the resonant module in conjunction with a Si-nanowire transistor to achieve a resonance frequency and $Q_{int}$ in excess of 2 GHz and 2000, respectively. We detect QD DRTs and QD-Boron ICTs and operate these as SEB and in-situ dispersive readout, respectively. We find a large SNR resulting in the lowest so far reported $t_\text{min}$ for dispersively detected charge sensors using a multi-module assembly and approaching the state-of-the-art $t_\text{min}$ of the monolithically integrated resonators and that of the SET. Finally, we use the large SNR to measure the charge noise of the ICT over 9 orders of magnitude up to 500 kHz using purely voltage spectroscopy. This work pushes the limits of charge sensitivity using dispersive readout and provides a transferable platform to improve the sensitivity of other charge-based quantum systems. The improvement in read-out sensitivity will directly impact the achievable read-out time and state fidelity of semiconductor spin qubit systems helping in the development of a fault-tolerant quantum computer.
\par


\section*{Acknowledgement}
This research was supported by European Union’s Horizon 2020 research and innovation programme under grant agreement no.\ 951852 (QLSI), and by the UK's Engineering and Physical Sciences Research Council (EPSRC) via the Cambridge NanoDTC (EP/L015978/1). F.E.v.H. acknowledges funding from the Gates Cambridge fellowship (Grant No. OPP1144). L.C. acknowledges funding from EPSRC Cambridge UP-CDT EP/L016567/1. J. W. A. R. acknowledges funding from the EPSRC Core-to-Core International Network Grant “Oxide Superspin” (No. EP/ P026311/1). M.F.G.Z. acknowledges a UKRI Future Leaders Fellowship [MR/V023284/1]. We thank David F. Wise for discussions about Bayesian inference fitting.

\appendix

\section{Design details of superconducting resonator and filter}\label{app:sec:resonator-layout}
\begin{figure}
    \centering
    \includegraphics{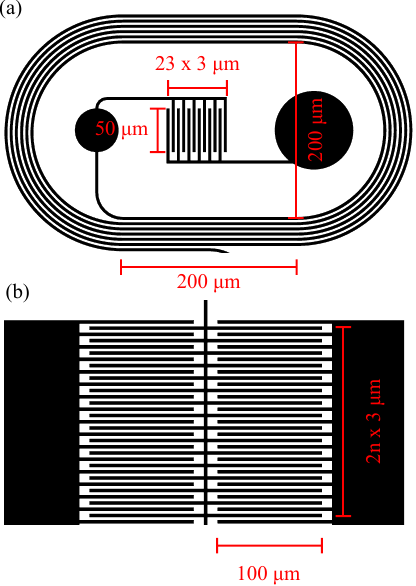}
    \caption{Schematic of spiral inductor and coupling capacitor (a) and section of the interdigitated filter capacitor (b) with dimensions.}
    \label{fig:app:resonator-design_detailed}
\end{figure}

In this section, we discuss the design of the inductance and capacitance values of the superconducting resonator and how these translate into its physical dimensions. Given an estimate of the parasitic capacitance on the load chip of the order of 200~fF, we targeted inductances on the order of 20-40~nH to achieve a $f_\text{r}\approx 2$~GHz. Given an estimate of parasitic losses of the circuit between 0.1 - 1~M$\Omega$ we calculated the required coupling capacitor for the matching condition [$C_{c,\text{matching}} \approx (L + \sqrt{L^2 + 4LCZ_0R})/ 2Z_0R$)] between 10-40~fF. \par

In order to estimate the inductance and capacitance of the various circuit elements, we used AWR Microwave Office, a finite element RF simulator, to simulate the elements before fabrication. We designed spiral inductors with an elongated middle section to create space for the coupling capacitor (see Fig.~\ref{fig:app:resonator-design_detailed}a). There are no simple, analytical equations for the inductance of such spiral inductors. Nevertheless, the inductance can be guided by equations for circular spiral inductors, namely $L \propto n^2 d$ where $n$ is the number of turns and $d$ is the diameter~\cite{mohan1999}. In our designs, we fixed the inner diameter to 200~$\mu$m with a horizontal elongation of 200~$\mu$m and used $n$ as a tuning parameter to select the inductance. \par

For the coupling capacitor and filter capacitor, we made use of interdigitated capacitors (Fig.~\ref{fig:app:resonator-design_detailed}a and b). The capacitance of these can be predicted using conformal mapping techniques \cite{igreja2004}. For design purposes, it is useful to note that the capacitance is approximately linear with the number and length of fingers of the capacitor. To keep the designs between resonators with different $C_c$ as similar as possible, we fixed the finger length to 50~$\mu$m and varied the number of fingers between 4 and 16 to achieve a capacitance of 10-40~fF. For the filter capacitor, we set the finger length to 100~$\mu$m instead and used 100 fingers on either side of the capacitor to get to a capacitance of $\approx$ 1~pF.

A final set of parameters are the track width ($w$) and separation ($s$). In general, lower values of $w$ and $s$ make the designs smaller leading to lower parasitic capacitances to ground and smaller footprints, however they can also make fabrication more challenging and increase capacitance between adjacent turns of the inductor. Finally, if there are impurities in the substrate these can couple more strongly with confined electric fields between features with smaller $s$ and therefore reduce the $Q$ of the resonator. For our designs, we chose 3~$\mu$m for $w$ and $s$.

\section{Resonator to device connection}\label{app:sec:device-image}

\begin{figure}[h]
    \centering
    \includegraphics{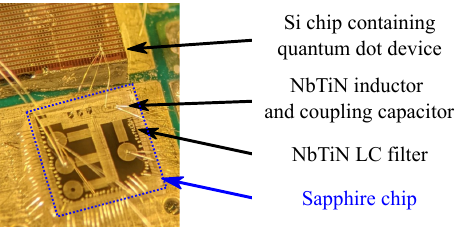}
    \caption{Picture showing resonator components wire-bonded to device. The bottom chip contains the NbTiN resonator and coupling capacitor (top right) and four LC filters on sapphire, three of them connected to the source-drain and gate of the device. The top chip contains the silicon QD device.}
    \label{fig:app:resonator-bonded}
\end{figure}

\section{Inductance correction for a finite filter capacitor}\label{sec:app:Cfilt-correction}
\begin{figure}
    \centering
    \includegraphics{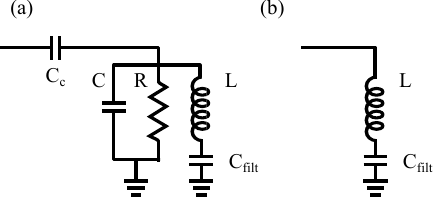}
    \caption{(a) Circuit schematic of capacitively coupled LCR resonator with filter capacitance. (b) Circuit schematic of inductor and filter capacitor. }
    \label{fig:app:Leff-calcs}
\end{figure}

For a capacitively coupled resonator with a filter capacitor behind the inductor (Fig.~\ref{fig:app:Leff-calcs}a), it can be shown that in the limit of the negligible resistances the resonance frequency, $\omega_r$, is given by

\begin{equation}\label{eq:app:fr_with_LC}
    \omega_r^2 = \frac{1}{L(C+C_c)} \left( 1+ \frac{C+C_c}{C_{filt}}\right),
\end{equation}
\noindent where $L$, $C_c$, $C$, $C_{filt}$ are the inductance, coupling capacitance, resonator capacitance and filter capacitance respectively. Noting that the resonance frequency without $C_{filt}$ is given by the first term, we can see that the resonance frequency is corrected upwards by a factor depending on how large the filter capacitance is with respect to the bare capacitance of the circuit. Intuitively, this can be understood by analysing the impedance of the inductor in series with the filter capacitor (Fig.~\ref{fig:app:Leff-calcs}b) and therefore calculating an effective inductance given by

\begin{eqnarray}
    Z_{LCfilt} = j\omega L + \frac{1}{j\omega C_{filt}} 
    = j\omega L \left(1- \frac{1}{\omega^2 LC_{filt}}\right) \\
    L_{eff} = L \left(1- \frac{1}{\omega^2 LC_{filt}}\right).
\end{eqnarray}

\par

Inserting Eq. \ref{eq:app:fr_with_LC} at resonance when $\omega = \omega_r$ this simplifies to:

\begin{equation}
    L_{\text{eff}} = L \left(1- \frac{C+C_c}{C_{filt}+C+C_c}\right) \approx  L \left(1- \frac{C+C_c}{C_{filt}}\right),
\end{equation}

\noindent where the approximation holds for $C_{filt} \gg C+C_c$. We have shown that $C_{filt}$ introduce a correction term to $\omega_r$ which can be understood as a reduction of the effective inductance of the resonator. 

\section{Large stability map}\label{app:sec:stability-map-large}

The stability map of the device under study shows two Boron-reservoir-transitions (BRT) intersected by several ICTs. DRTs can also be observed near the ICTs.

The different Boron acceptors and quantum dots can be distinguished by their different slope with respect to Vbg and Vg indicating different lever arms.

\begin{figure}[h]
    \centering\includegraphics{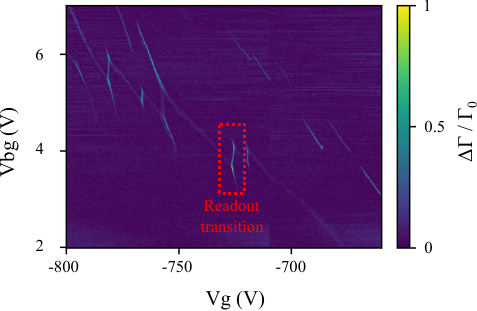}
    \caption{Stability map of gate voltage ($V_\text{g}$) against back-gate voltage ($V_\text{bg}$) recorded using reflectometry. The stability map shows two Boron-reservoir-transitions (the two diagonal lines with larger slope) intersected by several DRTs (quasi-vertical lines) creating ICTs at the intersection. The location of the readout transition investigated here is clearly marked.}
    \label{fig:app:stability-map-large}
\end{figure}

\section{Fitting dispersive shift}\label{sec:app:fitting-freq-shift}

\begin{figure}
\includegraphics{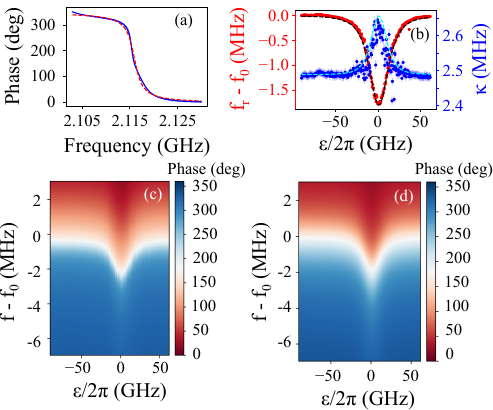}
\caption{\label{fig:app:fitting-freq-shift}Fitting procedure for dispersive shift measurements. (a) Example fit to the phase of the reflected signal at large detuning . (b) The frequency shift and $\kappa$ as a function of detuning for the DRT and ICT respectively. The dashed lines indicate the Bayesian fit extracted from fitting the full phase data. (c) Data and (d) fit of the whole ICT resonance using a charge qubit-resonator interaction model.}
\end{figure}

In the case of the ICT, the frequency shift can be modelled by a dispersive shift caused by a charge qubit onto the resonator described by the Hamiltonian \cite{Ibberson2021}:
\begin{equation}
    H = \hbar \Delta_0 a^\dagger a + \frac{\hbar\Delta}{2}\hat{\sigma_z} + \hbar g_\text{eff} (a\sigma_+ + a^\dagger \sigma_-),
\end{equation}
\noindent where $\Delta_0 = \omega_0 - \omega$ is the frequency detuning from the resonance frequency $\omega_0$, $\Delta = \Omega - \omega_0$ is the resonator-qubit detuning with the qubit frequency $\Omega$, $g_\text{eff}$ = $g_02t_c/\Omega$ is the effective charge coupling rate. The shift in resonance frequency and $\kappa^*$ are given by

\begin{eqnarray}
     (\omega_r - \omega_0) / 2\pi  = \frac{g_\text{eff}^2\Delta}{\Delta^2 + \gamma^2/4} \label{eq:app:ICT-fshift} \\
    \kappa^* = \kappa + \frac{g_\text{eff}^2\gamma}{\Delta^2 + \gamma^2/4}
\end{eqnarray}
\noindent where $\omega_r$  is the shifted resonance frequency as a function of detuning, $\kappa$ is the bare photon loss rate and and $\gamma$ is the charge qubit decoherence. \par

These equations allow us to fit the dispersive shift of the ICT on the resonator from the signal reflected by the resonator. Due to standing waves in the reflected magnitude which made fitting difficult, we instead fit the phase as this is known to give a reliable extraction of $\omega_r$ and $\kappa^*$ \cite{petersan1998}. The phase response of a resonator is given by:

\begin{equation}\label{eq:app:resonator-phase}
    \phi(f) = \phi_0 + 2\text{tan}^{-1}\left[\frac{2}{\kappa}(f-f_0)  \right]
\end{equation}

\noindent where $\phi_0$ is a constant phase offset.

Combining equations \ref{eq:app:ICT-fshift} to \ref{eq:app:resonator-phase}, 
we use Bayesian inference to extract the parameters of interest from the phase data. Each parameter is treated as a random variable with a prior distribution and Markov Chain Monte Carlo (MCMC) sampling is used along with the likelihood of the experimental data to sample from the posterior distribution of those parameters \cite{geyer1992,hoffman2014}. This provides us with an accurate estimate of the true uncertainty associated with these parameters based on our results \cite{barber2012,kinz-thompson2021,escalona2023}. Using this method, we find $g_0/2\pi=168\pm 0.7$~MHz, $2t_c= 17.7\pm 0.1$~GHz and $\gamma/2\pi=1.45 \pm 0.1$~GHz. We plot the phase data along with its fit in Fig. \ref{fig:app:fitting-freq-shift}c and d.

To benchmark the fit we additionally extract $\kappa^*$ and $f_r$ from line-cuts of the phase data at fixed detuning by fitting to equation \ref{eq:app:resonator-phase} (example fit see Fig. \ref{fig:app:fitting-freq-shift}a). The phase change of $\sim$ 360 degrees is characteristic of an overcoupled resonator. 
We plot the extracted $\kappa^*$ and $f_r$ as well as those predicted by the Bayesian inference to the full phase data in Fig. \ref{fig:app:fitting-freq-shift}b and find them to be in good agreement. There is significant noise in the extracted $\kappa^*$ which we attribute to charge noise causing discrete detuning shifts (this is also observable in the sharp lines in Fig. \ref{fig:app:fitting-freq-shift}c).

\section{Effect of charge noise on Fresnel blobs}\label{sec:app:Fresnel-chargenoise}

\begin{figure}
    \centering
    \includegraphics{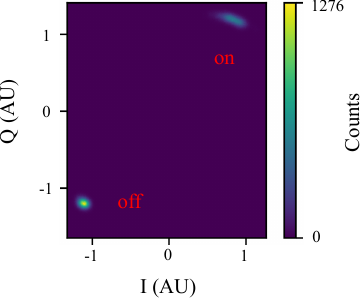}
    \caption{Fresnel blobs of DRT with $t_\text{meas}$ = 20~$\mu$s. Note that the on state is smeared by charge noise while the off state is not.}
    \label{fig:app:Fresnel-CN}
\end{figure}

In the high $t_\text{meas}$ regime of the SNR measurements (Fig. \ref{fig:SNR-tmin}b), we see a reduction in the rate of increase of the SNR. This can be attributed to a change in the noise characteristic of the measurement, changing the exponent in Eq.~\ref{eq:SNR_equation}. At low $t_\text{meas}$ the noise is white noise (cryogenic amplifier), leading to a noise power dependence of $N \propto$ 1/$t_\text{meas}$ arising from sampling from a normal distribution around a mean, and therefore a SNR $\propto$ $t_\text{meas}$. At large $t_\text{meas}$, excess noise manifest reducing the SNR. This can be demonstrated by plotting the Fresnel blobs of a measurement at large $t_\text{meas}$ (Fig. \ref{fig:app:Fresnel-CN}). Note how the on-state measurement, being detuning dependent, shows a significant elongation of the blob while the off-state blob remains mostly unaffected. \par

The two main sources of added noise are due to detuning ($\epsilon$) and tunnel noise ($\Gamma$ or $\Delta_c$). Changes in detuning shift the measurement point away from the centre of the peak which reduces the signal. As the slope of the signal with detuning is zero at $\epsilon = 0$ (see Fig.\ref{fig:Charge_noise} upper right insert), the signal is to first order insensitive to detuning noise. Tunnel noise can couple directly into the signal by changing the peak height, however is often less severe. Coincidentally, voltage spectroscopy is also sensitive to tunnel noise and therefore the difference in the noise characteristic from peak tracking (3.9~$\mu$eV$/\sqrt{\text{Hz}}$) and voltage spectroscopy (4.0~$\mu$eV$/\sqrt{\text{Hz}}$) can be used to demonstrate this.\par

\section{Kinetic inductance and losses introduced in resonator by large RF driving power}\label{sec:app:highpower-loss_lkin}

\begin{figure}
    \centering
    \includegraphics{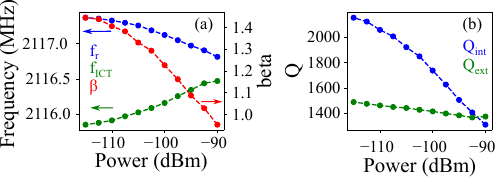}
    \caption{Effect of increasing power (a) on $f_\text{r}$ (blue), ICT shifted $f_\text{r}$ (green), $\beta$ and (b) $Q_{int}$ (blue), $Q_{ext}$ (green). Note that when $Q_{int}$ = $Q_{ext}$, $\beta$ = 1.}
    \label{fig:app:resonance-vs-power}
\end{figure}
At large driving powers, breaking of Cooper pairs can lead to an increased kinetic inductance as well as additional losses in the resonator. The kinetic inductance shifts the resonance frequency down, while the losses reduce $Q_{int}$ and therefore reduce the coupling constant $\beta$. \par
To analyse this effect we measure the frequency spectrum of the resonator near the ICT as a function of power. We plot $f_\text{r}$, $\beta$, $Q_{int}$ and $Q_{ext}$ in Fig. \ref{fig:app:resonance-vs-power}. Additionally, we also plot the $f_\text{r}$ shifted by the ICT which displays a reduced shift due to photon back-action. We find a decrease in $f_\text{r}$ by 0.56 MHz between -115 to -90~dBm of power which constitutes an inductance change of 16~$\pm$~3~pH assuming a purely inductive change.\par
We also find a monotonic decrease in $Q_{int}$ with power, affecting the impedance matching $\beta$. We find critical coupling $\beta$ = 1 at a power of -91.5 dBm. Note that this is significantly above the optimum read-out power found in this work.

\section{Aliasing effect of finite sampling on charge noise}\label{sec:app:Chargenoise-sampling}
When taking the Fourier transform of a sampled data set, duplicates of the Fourier spectrum are created in frequency space that are separated by the sample frequency $f_\text{s}$. In a non-ideal situation this will create an overlap between the true frequency spectrum and it's duplicates, an effect known as aliasing. This effect can be effectively avoided if the frequency spectrum of the data has a frequency cut-off, $f_\text{c}$, which is less or equal than half the $f_\text{s}$, known as the Nyquist condition, i.e. $f_\text{s} \leq 2 f_\text{c}$. Meeting the Nyquist condition can be effected by increasing the sample rate or using a low-pass filter. 

In Fig.~\ref{fig:Charge_noise}, we find an upwards bend in the charge noise data, in the range f $>$ 10$^{-1}$ for the peak tracking and the range f $>$ 10$^{5}$ for the voltage spectroscopy. We attribute this to an artefact caused by the aliasing effect described above. In principle this can be avoided by implementing a low-pass filter before the data is sampled, however depending on the frequency drop-off of the filter there may still be a finite contribution to the charge noise. The finite noise contribution can be modelled by a duplication of the charge noise offset by a frequency equal to the sampling frequency $f_\text{s}$:

\begin{equation}\label{eq:See-corrected}
    S_{\epsilon\epsilon} = S_0  \left( \frac{1}{f^a} + \frac{1}{(f_\text{s}-f)^a} \right)
\end{equation}
\par

In the case of our voltage-spectroscopy measurement we sampled at a frequency of 1~MHz, using a low-pass filter of 100~kHz with 6 dB drop-off per octave
(implemented by Stanford Research SR560 pre-amplifiers with variable low-pass filter), fulfilling the Nyquist condition. Because of the presence of this filter there is much less of an aliasing effect in the voltage-spectroscopy spectrum (compared to that of the peak-tracking) but some amount remains above 100~kHz. \par

For the peak-tracking, on the other hand, there is effectively no analogue filtering in the relevant frequency range. Peak positions are sampled at an $f_\text{s}$ of 10.4~Hz (given by the speed of the voltage ramp of 83.33~Hz over 8 averages) and the low-pass filter of the SR560 is set at 10~kHz. Reducing the low pass-filter frequency will interfere with the peak shape and therefore make accurate determination of the peak location impossible. This makes it difficult to remove any aliasing contribution for charge noise obtained using the peak-tracking method by analogue methods. Digital averaging and peak fitting introduces some amount of filtering above the averaging frequency and the ramp frequency respectively - however neither of these is able to fulfill the Nyquist condition since they are always limited to above the sampling frequency.

\bibliography{General}

\end{document}